\documentclass[prd,preprint
,superscriptaddress,showpacs,showkeys,nofootinbib
,tightenlines
]{revtex4}
\usepackage{mathrsfs}
\usepackage{latexsym,bm}
\usepackage{graphicx}
\usepackage{indentfirst}
\usepackage{slashed}
\usepackage{amsmath}
\usepackage{amssymb}
\usepackage{color}
\usepackage{hyperref}
\usepackage{epsfig}
\usepackage[titletoc]{appendix}
\usepackage{multirow}%
\usepackage{rotating}
\usepackage{epstopdf}
\usepackage{extarrows}
\usepackage{tabularx}
\usepackage{soul} 
\usepackage{xcolor}
\usepackage{graphicx}
\usepackage{lipsum}

\newcommand{\be}{\begin{equation}} \newcommand{\ee}{\end{equation}}
\newcommand{\ba}{\begin{array}{c}} \newcommand{\ea}{\end{array}}
\newcommand{\bea}{\begin{eqnarray}} \newcommand{\eea}{\end{eqnarray}}

\newcommand{\order}[1]{\mathcal{O}\left(#1\right)}
\newcommand{\md}{m_D}
\newcommand{\mx}{m_{D^\ast}}
\newcommand{\g}{g_0}
\newcommand{\add}{A_0(\md^2)}
\newcommand{\ads}{A_0(\mx^2)}
\newcommand{\bdd}{B_0(\md^2,0,\mx^2)}
\newcommand{\bds}{B_0(\mx^2,0,\md^2)}
\newcommand{\cds}{C_0(\mx^2,0,\md^2,0,\md^2,\mx^2)}
\newcommand{\al}{&\!\!\!\!}

\begin{document}{
\title{\Large  Study of open-charm {\boldmath$0^+$} states in unitarized chiral effective theory with one-loop potentials}
\author{Meng-Lin Du}
\email{du@hiskp.uni-bonn.de}
\affiliation{Helmholtz-Institut f\"ur Strahlen- und Kernphysik and
Bethe Center for Theoretical Physics, Universit\"at~Bonn, D--53115
Bonn, Germany}
\author{Feng-Kun Guo}
\email{fkguo@itp.ac.cn}
\affiliation{CAS Key Laboratory of Theoretical Physics,  Institute of
Theoretical Physics, Chinese~Academy~of~Sciences,
Beijing~100190, China} \affiliation{School of Physical Sciences, University of Chinese Academy of
Sciences,\\ Beijing 100049, China}
\author{Ulf-G. Mei{\ss}ner}
\email{meissner@hiskp.uni-bonn.de}
\affiliation{Helmholtz-Institut f\"ur Strahlen- und Kernphysik and
Bethe Center for Theoretical Physics, Universit\"at~Bonn, D--53115
Bonn, Germany}
\affiliation{Institute for Advanced Simulation, Institut f{\"u}r
Kernphysik and J\"ulich Center for Hadron Physics, Forschungszentrum~
J{\"u}lich, D-52425 J{\"u}lich, Germany}
\author{ De-Liang Yao}
\email{deliang.yao@ific.uv.es}
\affiliation{Institute for Advanced Simulation, Institut f{\"u}r
Kernphysik and J\"ulich Center for Hadron Physics, Forschungszentrum~
J{\"u}lich, D-52425 J{\"u}lich, Germany}
\affiliation{Instituto de F\'{\i}sica Corpuscular (centro mixto CSIC-UV), Institutos de Investigaci\'on de Paterna,
Apartado 22085, 46071, Valencia, Spain}

\begin{abstract}

Chiral potentials are derived for the interactions between Goldstone bosons and
pseudoscalar charmed mesons up to next-to-next-to-leading order in a covariant
chiral effective field theory with explicit vector charmed-meson degrees of
freedom. Using the extended-on-mass-shell scheme, we demonstrate that the
ultraviolet divergences and the so-called power counting breaking terms can be
properly absorbed by the low-energy constants of the chiral Lagrangians. We
calculate the scattering lengths by unitarizing the one-loop potentials and
fit them to the data extracted from lattice QCD. The obtained results are
compared to the ones  without an explicit contribution of vector charmed mesons
given previously. It is found that the difference is negligible for $S$-wave
scattering in the threshold region. This validates the use of $D^\ast$-less
one-loop potentials in the study of the pertinent scattering lengths. 
We search for dynamically generated open-charm states with
$J^P=0^+$ as poles of the $S$-matrix on various Riemann sheets. The
trajectories of those poles for varying pion masses are presented as well.

\end{abstract}

\maketitle

\section{Introduction}

In the past two decades, many excited charmed states have been observed
experimentally~\cite{Bevan:2014iga,Aubert:2003fg,Krokovny:2003zq,Besson:2003cp}
and further  experiments are intended either to investigate their
properties more precisely or to search for more new states, e.g., by the LHCb
Collaboration~\cite{Aaij:2015vea}. The conventional quark-potential models
provide a successful description of most of those low-lying excitations, see
Ref.~\cite{Chen:2016spr} for a recent review. However, quantum 
chromodynamics (QCD) at low energies has a much richer structure than quark
models. There exist observed charmed mesons whose properties are in
disagreement with the expectations from quark models, of which the most
interesting one is the $D_{s0}^*(2317)$. It was first observed by the BABAR
Collaboration in the inclusive $D_s^+\pi^0$ invariant mass distribution and
later confirmed by Belle and CLEO
Collaborations~\cite{Aubert:2003fg,Krokovny:2003zq,Besson:2003cp}. It couples to
the $DK$ channel and decays mainly into the isospin breaking channel $D_s\pi$
due to its location below the $DK$ threshold.  Many theoretical investigations
were triggered consequently, attempting to reveal the nature of the $D_{s0}^*(2317)$
as well as other newly observed charmed states with $J^P=0^+$ and trying to
reveal their internal structure.  For instance, the $D_{s0}^*(2317)$ has been
suggested to be a $DK$ bound state~\cite{Barnes:2003dj}. Were this
interpretation true, one can learn much about the interaction between the 
$D/D_s$
mesons and $\pi/K$ mesons from studying the $D_{s0}^*(2317)$ and related states.
This path has been followed in Refs.~\cite{Kolomeitsev:2003ac,
Guo:2006fu,Gamermann:2006nm,Hofmann:2003je,Guo:2008gp,Guo:2009ct,Cleven:2010aw,
Cleven:2014oka} where the $S$-wave interaction between charmed $D$ mesons and
Goldstone bosons (denoted as $\phi$ hereafter) has been studied
systematically up to the next-to-leading order (NLO) using chiral perturbation
theory (ChPT) for heavy mesons~\cite{Burdman:1992gh,Wise:1992hn,Yan:1992gz} in
combination with a unitarization procedure such as the one in
Ref.~\cite{Oller:2000fj}.

In the meantime, significant progress has also been made in lattice
QCD~\cite{Moir:2013ub,Cichy:2015tma}.
Using the L{\"u}scher formalism and its extension to coupled channels (for early works on this
topic, see e.g. Refs.~\cite{Liu:2005kr,Lage:2009zv}),
scattering lengths and recently phase shifts for the $D\phi$ interaction have
been calculated at unphysical quark
masses~\cite{Liu:2008rza,Liu:2012zya,Mohler:2012na,Mohler:2013rwa,
Lang:2014yfa,Moir:2016srx}. The first calculation only concerns the channels
free of disconnected Wick contractions~\cite{Liu:2008rza,Liu:2012zya},
i.e., $D\pi$ with isospin $I=3/2$, $D\bar{K}$ with $I=0,1$, $D_sK$ and $D_s\pi$. The
channels with disconnected Wick contractions
such as $D\pi$ with $I=1/2$ and $DK$ with $I=0$ were calculated later in
Refs.~\cite{Mohler:2012na,Mohler:2013rwa,Lang:2014yfa}.
On the one hand, the lattice results can be used to determine the low-energy
constants~(LECs) in the chiral
Lagrangian~\cite{Wang:2012bu,Liu:2012zya,Altenbuchinger:2013vwa,Yao:2015qia,
Guo:2015dha}.
On the other hand, with these lattice calculations more insights into the
nature of the $D_{s0}^*(2317)$ and other positive-parity charmed mesons are
obtained. In particular, in Ref.~\cite{Liu:2012zya}, it is concluded that the
lattice calculation of other channels performed there supports the interpretation that
$D_{s0}^*(2317)$ is dominantly a $DK$ hadronic molecule. In addition, using the
parameters fixed in that work, energy levels in the $I=1/2$ channel were
computed in Ref.~\cite{Albaladejo:2016lbb}, and a remarkable agreement 
with the lattice results reported in Ref.~\cite{Moir:2016srx} was found. 
This agreement was taken to be
as a strong evidence that the particle listed as $D_0^*(2400)$ in the Review of
Particle Physics~\cite{Olive:2016xmw} in fact corresponds to two states with
poles located at $\left(2105^{+6}_{-8}-i\,102^{+10}_{-12}\right)$~MeV and
$\left(2451^{+36}_{-26}-i\,134^{+7}_{-8}\right)$~MeV,
respectively~\cite{Albaladejo:2016lbb}, 
similar to the well-known two-pole scenario of the $\Lambda(1405)$~\cite{Oller:2000fj}.
In this scenario, the puzzle that the
non-strange $D_0^*(2400)$ has a mass larger than the strange partner
$D_{s0}^*(2317)$ can be easily understood. The poles were searched for in
unitarized ChPT with the interaction kernel computed at NLO. In view of the
phenomenological importance of the $D_{s0}^*(2317)$ and $D_0^*(2400)$, it is
crucial to check the stability of the NLO predictions by extending to the
next-to-next-to-leading order (NNLO), which is one of the purposes of this
work.

When massive matter fields are included in ChPT, the nonvanishing matter-field
mass in the chiral limit leads to the notable power
counting breaking (PCB) issue~\cite{Gasser:1987rb}: all loop graphs containing
internal matter field propagators start contributing at
$\order{p^2}$.\footnote{The closed matter field loops are not taken into
account since they are real below the two-matter-field threshold and can
be absorbed by the redefinition of LECs \cite{Gasser:1987rb}.} Various
approaches have been proposed to remedy the PCB issue,
e.g. heavy baryon ChPT~\cite{Jenkins:1990jv,Bernard:1992qa}, 
infrared regularization~\cite{Becher:1999he}, and the
extended-on-mass-shell (EOMS) scheme~\cite{Fuchs:2003qc}. Recently, the EOMS
scheme has been demonstrated to be a good solution to the PCB problem. It has
been successfully applied to the study of $\pi N$ scattering up to
$\mathcal{O}(p^3)$~\cite{Alarcon:2012kn} and
$\mathcal{O}(p^4)$~\cite{Chen:2012nx}, and  up to leading
one-loop order in the presence of $\Delta$-resonance~\cite{Yao:2016vbz}.

The first aim of this paper is to present a full calculation of $D\phi$
scattering using the EOMS scheme within a manifestly Lorentz invariant chiral
effective theory  with explicit vector charmed mesons, to be denoted as $D^*$, 
up to NNLO, i.e. the leading one-loop order.
The first study on $D\phi$ scattering to one-loop was made in
Ref.~\cite{Liu:2009uz} in the framework of non-relativistic heavy meson ChPT,
which neglects sizeable recoil corrections~\cite{Geng:2010vw}.\footnote{
Such recoil corrections can be restored by using the extended heavy baryon propagator
$i/(v\cdot k + k^2/2m)$  instead of $i/v\cdot k$~\cite{Bernard:1993ry}.}  The
first one-loop calculation within the covariant formalism was given in
Ref.~\cite{Geng:2010vw}. However, in that paper the NNLO contact terms of $D\phi$ are
not included and the kinetic term for the $D^\ast$ is incomplete, both of which
are necessary for a proper renormalization in the EOMS scheme.
Furthermore, the scattering amplitudes are calculated perturbatively without
considering resonant charmed mesons close-to or even below thresholds such as
the $D_{s0}^*(2317)$ in the channel $(S,I)=(1,0)$, where $S$ and $I$
denote the strangeness and isospin, respectively. Thus, the results in
those channels of such a calculation are incomplete and thus can not be 
considered  
significant.
In our previous work~\cite{Yao:2015qia}, the $D\phi$ scattering amplitudes
are presented up to NNLO explicitly in the absence of the $D^*$. Then, the
scattering lengths are calculated based on the unitarized amplitudes, which are
also used to fit to the lattice QCD data at unphysical pion masses.
In order to judge the importance of the $D^*$, a selection of all the diagrams
containing the $D^*$ is calculated and found to be negligible. However, a
complete calculation including the $D^\ast$ mesons is still lacking and,
furthermore, a systematic renormalization using the EOMS scheme is required. It is
also important to check whether the full finite $c$-quark mass effects,
corresponding to including the $D^*$ which are degenerate with the $D$ mesons
in the heavy quark limit, are sizeable. These gaps will be closed in this paper.

Specifically, in Section~\ref{sec:theo}, we derive the covariant one-loop
$D\phi$ scattering potentials  with the $D^*$ resonances as dynamical degrees
of freedom. We perform renormalization in the EOMS scheme and explicitly show
that the ultraviolet~(UV) divergences  and PCB terms can be absorbed by redefining
the LECs. The $D^\ast$-less case has been accomplished formally using the path
integral formalism in Refs.~\cite{Du:2016ntw,Du:2016xbh}.
Then a unitarization procedure is taken  to
generate resonances not far from the corresponding thresholds.
In such a scheme, one can also deal with larger pion masses as compared
with a purely perturbative approach. However, in general crossing symmetry
is no longer exactly fulfilled.
When the unitarization is extended to the one-loop order, an additional
subtraction in the potentials is needed to remove the right-hand cut in the
$N(s)$ functions, see below~\cite{Oller:2000fj,Guo:2011pa,Yao:2015qia}. In
Section~\ref{sec:num}, by fitting the so-obtained scattering lengths to the
lattice results, we determine
the LECs in the effective Lagrangian. Then we search for poles in the
unitarized amplitudes, and study their trajectories with varying pion
mass. Section~\ref{sec:sum} comprises a brief summary. The explicit
UV-part and EOMS subtractions of the LECs are collected in Appendix~\ref{sec:app}.

\section{Theoretical discussions of the $D\phi$ interactions}\label{sec:theo}

\subsection{Effective Lagrangian}

To set up the  effective Lagrangian, we first specify the corresponding power
counting rules. At low energies, the external momenta as well as the
masses of the Goldstone bosons are counted as $\order{p}$. However, the
nonvanishing masses of the $D$ and $D^*$ in the chiral limit introduce new
scales $M_0$ and $M_0^*$, both counted as $\order{1}$. As a result, at low
energies, the temporal components of the momenta of the $D$ and $D^*$ are
counted as $\order{1}$, while the spatial components are counted as
$\order{p}$. Therefore, the virtuality $q^2-M_0^{(*)2}$ in the
propagators scales as $\order{p}$, and the propagators scale as
$\order{p^{-1}}$. The Goldstone boson propagators are counted as
$\order{p^{-2}}$ as usual. Based on the counting rules for the vertices and
propagators, one can assign a chiral order for a given Feynman diagram, and thus
for any physical quantity. However, for the specific Feynman graphs with loops,
there exist terms with chiral order lower than the naive power counting order,
which are called PCB terms. In the EOMS scheme, the PCB terms are
absorbed into the redefinition of the LECs so that the resulting physical
observables obey the power counting rules.

The effective Lagrangian relevant to our calculation of the $D\phi$ potentials
up to leading one-loop order can be written as
\bea
\mathscr{L}_{\rm eff}=\sum_{i=1}^2\mathscr{L}_{\phi\phi}^{(2i)}+\sum_{j=1}^3\mathscr{L}_{D\phi}^{(j)}+\sum_{k=1}^2\mathscr{L}_{D^\ast\phi}^{(k)}+\sum_{l=1}^3\mathscr{L}_{D^\ast D\phi}^{(l)}\
\eea 
with the superscripts specifying the chiral dimension. The needed terms in
the Goldstone sector read~\cite{Gasser:1984gg} 
\bea
\mathscr{L}_{\phi\phi}^{(2)}&=&\frac{F_0^2}{4} \left\langle \partial_\mu U
(\partial^\mu U)^\dagger \right\rangle + \frac{F_0^2}{4}\left\langle\chi
U^\dagger+ U\chi^\dagger\right\rangle\ ,\nonumber\\
\mathscr{L}_{\phi\phi}^{(4)}&=& L_4 \left\langle\partial_\mu U(\partial^\mu
U)^\dagger \right\rangle \left\langle\chi U^\dagger+U\chi^\dagger \right\rangle+
L_5 \left\langle\partial_\mu U(\partial^\mu U)^\dagger \left(\chi
U^\dagger+U\chi^\dagger\right) \right\rangle +\ldots\ .
\eea
where the trace in flavor space is denoted by $\langle\cdots\rangle$, $F_0$ is
the pion decay constant in the chiral limit, and $L_{4,5}$ are
LECs. Furthermore, $\chi=2B_0\,{\rm diag}(m_u,m_d, m_s)$,  with $B_0$ a
constant
related to the quark condensate, and $U=\exp\left({i\sqrt{2}\phi}/{F_0}\right)$,
with
\bea
\phi=\begin{pmatrix}
   \frac{1}{\sqrt{2}}\pi^0 +\frac{1}{\sqrt{6}}\eta  & \pi^+ &K^+  \\
     \pi^- &  -\frac{1}{\sqrt{2}}\pi^0 +\frac{1}{\sqrt{6}}\eta&K^0 \\
     K^-&\bar{K}^0&-\frac{2}{\sqrt{6}}\eta
\end{pmatrix} .
\eea
The terms corresponding to interactions between the $D=(D^0,D^+,D_s^+)$ mesons and 
the Goldstone bosons are given by~\cite{Burdman:1992gh,Wise:1992hn,Yan:1992gz,Guo:2008gp,Yao:2015qia}
\footnote{As discussed in Ref.~\cite{Du:2016ntw}, the dimension-three terms proportional to
$g_{4,5}$  in that  work do not contribute to
the $D\phi$ scattering because their contribution from contact term to 
amplitudes will be canceled out by their contribution to
the wave function renormalization. }
\bea
\mathscr{L}^{(1)}_{D\phi}\al=\al\mathcal{D}_\mu D \mathcal{D}^\mu D^\dagger-{M}_0^2D
D^\dagger\ ,\nonumber\\
\mathscr{L}^{(2)}_{D\phi} \al=\al D\left(-h_0\langle\chi_+\rangle-h_1{\chi}_+
+ h_2\langle u_\mu u^\mu\rangle-h_3u_\mu u^\mu\right) {D}^\dag \nonumber\\
\al\al + \mathcal{D}_\mu D\left({h_4}\langle u_\mu
u^\nu\rangle-{h_5}\{u^\mu,u^\nu\}\right)\mathcal{D}_\nu {D}^\dag\ ,\nonumber\\
\mathscr{L}^{(3)}_{D\phi} \al=\al
D\biggl[ i\,{g_1}[{\chi}_-,u_\nu] +
{g_2}\left([u_\mu,[\mathcal{D}_\nu,u^\mu]] + [u_\mu,[\mathcal{D}^\mu,u_\nu]]
\right)\biggr]\mathcal{D}^\nu {D}^\dag \nonumber\\
\al\al + g_3
D\,[u_\mu,[\mathcal{D}_\nu,u_\rho]] \mathcal{D}^{\mu\nu\rho} {D}^\dag\ +h.c. \ ,
\eea
where $h_i$ and $g_j$ are LECs and the chiral building blocks are given by
\bea
u_\mu=i\bigg[u^\dagger\partial_\mu u-u\partial_\mu u^\dagger\bigg]\ ,\qquad U=u^2\ ,\qquad
\chi^\pm=u^\dagger\chi u^\dagger\pm u\chi^\dagger u\ .
\eea
The covariant derivative is defined via
\bea\label{eq:cov}
\mathcal{D}_\mu H=H(\overset{\leftarrow}{\partial_\mu}+\Gamma_\mu^\dagger)\ ,
\qquad \mathcal{D}_\mu H^\dagger=(\partial_\mu+\Gamma_\mu)H^\dagger\ ,
\eea
and $\mathcal{D}^{\mu\nu\rho}=\{\mathcal{D}_\mu,
\{\mathcal{D}_\nu,\mathcal{D}_\rho\}\}$, where $H\in\{D,D^\ast\}$ with
$D^\ast=(D^{\ast0},D^{\ast+},D_s^{\ast+})$. The so-called chiral connection in the covariant derivatives is defined as
$
\Gamma_\mu=\left(u^\dagger\partial_\mu u+u\partial_\mu
u^\dagger\right )/2.
$
Similarly, the relevant terms for the interaction between the $D^\ast$ and the Goldstone bosons
 are~\cite{Burdman:1992gh,Wise:1992hn,Yan:1992gz}
\bea
\mathscr{L}_{D^\ast\phi}^{(1)}\al=\al -\frac{1}{2}\mathcal{F}^{\mu\nu}\mathcal{F}_{\mu\nu}^\dagger+M_0^{\ast 2} D^{\ast\nu} D^{\ast\dagger}_\nu\ ,\nonumber\\
\mathscr{L}_{D^\ast\phi}^{(2)}\al=\al D_\mu^\ast\left[\tilde{h}_0\langle\chi_+\rangle+\tilde{h}_1{\chi}_+\right]D^{\mu\ast\dagger}\ ,
\eea
with $\tilde{h}_{0,1}$ analogous to $h_{0,1}$ and
$\mathcal{F}_{\mu\nu}=(\mathcal{D}_\mu D^\ast_{\nu}-\mathcal{D}_\nu
D^\ast_{\mu})$.
Finally, the LO axial coupling has the form
\bea
\mathscr{L}_{D^\ast D\phi}^{(1)}=i\,{g_0}\left(D^\ast_{\mu}u^\mu D^\dagger-D\,u^\mu D^{\ast\dagger}_\mu\right)\ .
\eea
As pointed out in Refs.~\cite{Krebs:2009bf,Yao:2016vbz}, the resonance-exchange
contributions of $\mathcal{O}(p^2)$ and $\mathcal{O}(p^3)$ can be taken into
account by shifting the coupling in the LO resonance-exchange contribution and
the LECs in the contact terms. This also holds true for our case. Thus,
we do not need the $\order{p^2}$ and $\order{p^3}$ terms for the $D^\ast
D\phi$ coupling.

 \begin{figure}[t]
\begin{center}
\includegraphics[width=\textwidth]{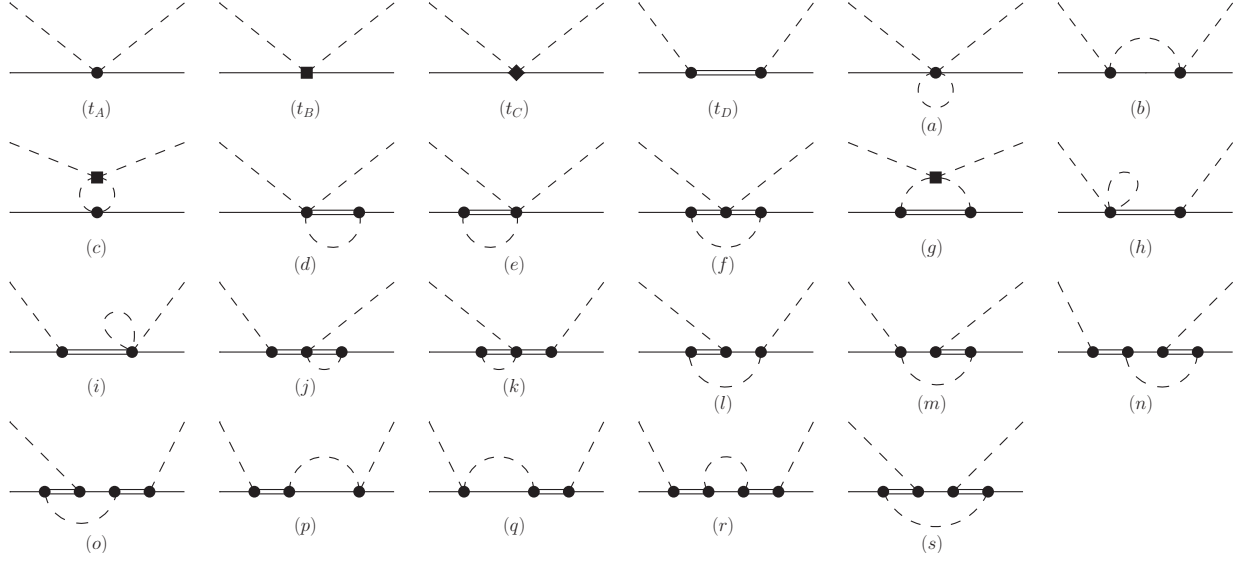}
\end{center}
\caption{Feynman diagrams contributing to $D\phi$ scattering up to NNLO with explicit $D^\ast$ mesons. The dashed, solid and double-solid lines stand for Goldstone bosons $\phi$, pseudo-scalar $D$ mesons and vector $D^\ast$ mesons, respectively. The dot, square and diamond represent vertices coming from Lagrangians of $\mathcal{O}(p^1)$, $\mathcal{O}(p^2)$ and $\mathcal{O}(p^3)$, in order.}
\label{fig:FD}
\end{figure}

\subsection{Chiral potentials up to leading one-loop order}

Up to NNLO, the Feynman diagrams needed for our calculation are displayed in
Fig.~\ref{fig:FD}. Accordingly, the chiral potential for the process
$D_1(p_1)\phi_1(p_2)\to D_2(p_3)\phi_2(p_4)$  can be written as
\bea\label{eq:potall}
\mathcal{V}_{D_1\phi_1\to D_2\phi_2}(s,t)=\mathcal{V}^{\rm(WT)}_{\rm LO}+\mathcal{V}^{\rm(EX)}_{\rm LO}+\mathcal{V}^{\rm(CT)}_{\rm NLO}+\mathcal{V}^{\rm(CT)}_{\rm NNLO}+\mathcal{V}^{\rm(Loop)}_{\rm NNLO}\ .
\eea
As usual, the Mandelstam variables are defined by $s=(p_1+p_2)^2$ and 
$t=(p_1-p_3)^2$, while $u$ can be obtained via 
$u=\sum_{i=1}^2(M_{D_i}^2+M_{\phi_i}^2)-s-t$.
The potentials at tree-level are given by
\bea
\mathcal{V}^{\rm(WT)}_{\rm LO}(s,t)\al=\al\mathcal{C}_\text{LO}\frac{s-u}{4F_0^2}\ ,\\
\mathcal{V}^{\rm(EX)}_{\rm LO}(s,t)\al=\al\mathcal{C}_S\frac{g_0^2}{F_0^2}\mathcal{F}_S(s,t)+\mathcal{C}_U\frac{g_0^2}{F_0^2}\mathcal{F}_U(s,t)\ ,\\
\mathcal{V}^{\rm(CT)}_{\rm NLO}(s,t)\al=\al\frac{1}{F_0^2}\bigg[-4h_0\mathcal{C}_0^{(2)}+
{2}h_1\mathcal{C}_1^{(2)}-2\mathcal{C}_{24}^{(2)}\mathcal{H}_{24}(s,t)+
2\mathcal{C}_{35}^{(2)}\mathcal{H}_{35}(s,t)\bigg]\ ,\\
\mathcal{V}^{\rm(CT)}_{\rm NNLO}(s,t) \al=\al
\frac{4g_1}{F_0^2}\bigg[\mathcal{C}_{1a}^{(3)}(p_1+p_3)\cdot(p_2+p_4)+
\mathcal{C}_{1b}^{(3)}(p_1+p_3)\cdot p_2\bigg]+
\frac{4\mathcal{C}_{23}^{(3)}\mathcal{G}_{23}(s,t)}{F_0^2},
\eea
where the involved coefficients corresponding to various scattering processes
are shown in Table~\ref{TabCoes1}. The functions in the $D^*$-exchange
potentials read
\bea
\mathcal{F}_S(s,t)\al=\al\frac{(p_1+p_2)\cdot p_4(p_1+p_2)\cdot p_2-M_0^{*2}p_2\cdot p_4}{M_0^{*2}(s-M_0^{*2})}\ ,\\
\mathcal{F}_U(s,t)\al=\al\frac{(p_1-p_4)\cdot p_2(p_1-p_4)\cdot p_4-M_0^{*2}p_2\cdot p_4}{M_0^{*2}(u-M_0^{*2})}\ .
\eea

\begin{table}
    \centering
    \caption{The coefficients in the tree-level amplitudes for the ten
relevant physical processes, with $\Delta_{K\pi}=M_K^2-M_\pi^2$.}\label{TabCoes1}
   \begin{tabular}{ll|c|cc|cccc|ccc}
   \hline\hline
   &
Physical~process &$\mathcal{C}_\text{LO}$&$\mathcal{C}_S$ & $\mathcal{C}_U$&$\mathcal{C}_{0}^{(2)}$&
$\mathcal{C}_{1}^{(2)}$&$\mathcal{C}_{24}^{(2)}$&$\mathcal{C}_{35}^{(2)}$&$\mathcal{C}_{1a}^{(3)}$&$\mathcal{C}_{1b}^{(3)}$&
$\mathcal{C}_{23}^{(3)}$\\
\hline
1&$D^0K^-\to D^0K^-$&1& 0 & 2 & $M_K^2$ & $-M_K^2$ & 1 & 1&  $M_K^2$ & 0 & 1 \\
2&$D^+K^+\to D^+K^+$&0&0 & 0 & $M_K^2$ & 0 & 1 & 0 &  0 & 0 & 0\\
3&$D^+\pi^+\to D^+\pi^+ $&1&0 & 2& $M_{\pi }^2$ & $-M_{\pi }^2$ & 1 & 1&  $M_{\pi }^2$ & 0 & 1\\
4&$D^+\eta\to D^+\eta $&0& $\frac{1}{3}$ & $\frac{1}{3}$ & $M_\eta^2$ & $-\frac{1}{3}M_{\pi }^2$ & 1 & $\frac{1}{3}$&  0 & 0 & 0\\
5&$D_s^+K^+\to D_s^+K^+ $&1& 0 & 2& $M_K^2$ & $-M_K^2$ & 1 & 1 &  $M_K^2$ & 0 & 1\\
6&$D_s^+\eta\to D_s^+\eta $&0& $\frac{4}{3}$ & $\frac{4}{3}$ &$M_\eta^2$ &  $\frac{4}{3}(M_\pi^2-2 M_{K
}^2)$   & 1 & $\frac{4}{3}$&  0 & 0 & 0 \\
7&$D_s^+\pi^0\to D_s^+\pi^0$ &0& 0 & 0 & $M_{\pi }^2$ & 0 & 1 & 0 &  0 & 0 & 0 \\
8&$D^0\eta\to D^0\pi^0$ &0& $\frac{1}{\sqrt{3}}$ & $\frac{1}{\sqrt{3}}$& 0&$-\frac{1}{\sqrt{3}} M_{\pi }^2$ & 0 & $\frac{1}{\sqrt{3}}$ &  0 & 0 & 0 \\
9&$D_s^+K^-\to D^0\pi^0$ &$-\frac{1}{\sqrt{2}}$& $\sqrt{2}$ & 0 & 0 &  $-\frac{1}{2
\sqrt{2}}(M_K^2+M_{\pi }^2 )$ & 0 & $\frac{1}{\sqrt{2}}$ &$-\frac{1}{\sqrt{2}}M_K^2$&$\frac{1}{\sqrt{2}}\Delta_{K\pi}$&$-\frac{1}{\sqrt{2}}$\\
10&$D_s^+K^-\to D^0\eta$ &$-\sqrt{\frac{3}{2}}$& $\sqrt{\frac{2}{3} }$ & $-\sqrt{ \frac{8}{3} }$ & 0 &  $\frac{1}{2\sqrt{6}}(5 M_K^2-3 M_{\pi }^2)$
  & 0 & $-\frac{1}{\sqrt{6}}$&$-\sqrt{\frac{3}{2}}M_K^2$&$\frac{-1}{\sqrt{6}}\Delta_{K\pi}$&$-\sqrt{\frac{3}{2}}$\\
\hline\hline
\end{tabular}
\end{table}

The functions in the NLO potentials read
\bea
\mathcal{H}_{24}(s,t,u)\al=\al 2h_2\,p_2\cdot p_4+h_4\,(p_1\cdot p_2 p_3\cdot p_4+p_1\cdot
p_4 p_2\cdot p_3)\ ,\\
\mathcal{H}_{35}(s,t,u)\al=\al h_3\,p_2\cdot p_4+h_5\,(p_1\cdot p_2 p_3\cdot p_4+p_1\cdot
p_4 p_2\cdot p_3)\ ,
\eea
while the one in the NNLO potentials is
\bea
\mathcal{G}_{23}(s,t,u) \al=\al -g_2\,p_2\cdot p_4(p_1+p_3)\cdot(p_2+p_4) \nonumber\\
\al \al + 2g_3\left[ (p_1\cdot p_2)( p_1\cdot p_4) p_1\cdot(p_2+p_4)
+(p_1\to p_3)\right]\ .
\eea
As for the one-loop potentials at NNLO,  the parts without explicit
$D^\ast$ mesons can be found in the appendix of Ref.~\cite{Yao:2015qia} and the
ones involving explicit $D^\ast$ states are too lengthy to be shown here. Note that 
$\mathcal{V}^{\rm(Loop)}_{\rm NNLO}$ in Eq.~(\ref{eq:potall}) contains the contribution from wave 
function renormalization as well. We
performed renormalization of the one-loop potentials using the so-called EOMS
scheme. In this scheme, the UV divergence are absorbed by the counterterms when
the bare LECs are expressed in terms of the renormalized ones via
\bea\label{eq:UVshift}
M_0^2 \al= \al M_0^{r 2}(\mu)+ \beta_{M_0^2} \frac{R}{16\pi^2 F_0^2}\ , \nonumber \\
M_0^{\ast 2} \al =\al M_0^{\ast r 2}(\mu)+ \beta_{M_0^{\ast 2}} \frac{R}{16\pi^2 F_0^2}\ , \nonumber \\
h_i\al=\al h_i^r(\mu)+\beta_{h_i}\frac{R}{16\pi^2F_0^2}\ ,\quad(i=0,1,\cdots,5)\nonumber\\
 g_j\al=\al g_j^r(\mu)+\beta_{g_j}\frac{R}{16\pi^2F_0^2}\ ,\quad(j=0,1,2,3)\ ,
\eea
where $R=\frac{2}{d-4}+\gamma_E-1-\ln(4\pi)$, with $\gamma_E$
the Euler constant and $d$ the number of  space-time dimensions. The $\beta$-functions are given in Appendix~\ref{sec:beta}. Here, $\mu$ is the scale introduced in dimensional regularization. Then additional subtractions are performed by splitting the UV-renormalized LECs via
\bea\label{eq:finiteshift}
h_i^r(\mu)\al=\al\bar{h}_i+\frac{\bar{\beta}_{h_i}}{16\pi^2F_0^2}\ ,\qquad (i=1,2,\cdots,5)\ ,\nonumber\\
g_0^{r}(\mu)\al=\al\bar{g}_0+\frac{\bar{\beta}_{g_0}}{16\pi^2F_0^2}\ ,
\eea
such that the PCB terms from the one-loop potentials are
canceled. The remaining LECs $g_1$, $g_2$ and $g_3$ are untouched at the chiral
order we are working. The coefficients $\bar{\beta}_{h_i}$ and
$\bar{\beta}_{g_0}$  are given in Appendix~\ref{sec:finite}.

\subsection{Partial waves and unitarization}

In the present paper, we do not consider the effect of isospin violation. It is convenient to study the potentials in the isospin basis instead of the
particle basis. All possible processes with definite strangeness $S$ and isospin $I$ can be obtained from the ten processes given in Table~\ref{TabCoes1}
by crossing  and isospin symmetry, see Refs.~\cite{Guo:2009ct,Yao:2015qia} for details.

Since the standard ChPT is organized
in a double expansion in terms of small external momenta and light quark masses,
it is expected to work well in the low-energy region.
With  increasing energy the convergence of the chiral series  becomes worse.
Especially, when the energy reaches the region where resonances appear, the
perturbative chiral potentials start to violate unitarity largely and
cannot be directly applied anymore. One way to restore unitarity is to
unitarize the potentials, but usually at the price of violating the crossing
symmetry. While the unitarity and analyticity of the single-channel $\pi \pi$
potentials are strictly restored within a range of
energies~\cite{Pelaez:2015qba}, a rigorous solution for the coupled-channel
case is still missing. A convenient
approximation is to treat the right-hand cut nonperturbatively, while the
cross-channel effects are incorporated in a perturbative
manner~\cite{Oller:1997ti,Oller:1998hw}.\footnote{A method of 
calculating the
left-hand cut nonperturbatively was proposed very recently in
Ref.~\cite{Entem:2016ipb}.} The unitarization is equivalent to a resummation of
the $s$-channel potentials, and can extend the applicable energy range of the
perturbative amplitudes. For instance,  the scattering data for the pion-kaon
systems up to $1.2$~GeV can be well
described~\cite{Oller:1997ti,Oller:1998hw,Oller:1998zr}.

Before unitarization, the partial wave projection to a definite orbital angular
momentum $l$ should be performed
\bea
\mathcal{V}_{l}^{(S,I)}(s)_{D_1\phi_1\to D_2\phi_2}
= \frac{1}{2}\int_{-1}^{+1}{\rm d}\cos\theta\,P_l(\cos\theta)\, \mathcal{V}^{(S,I)}_{D_1\phi_1\to
D_2\phi_2}(s,t(s,\cos\theta))\ ,
\label{eq:pwp}
\eea
where $\theta$ is the scattering angle between the incoming and outgoing particles in the center-of-mass frame, and the Mandelstam variable $t$
is expressed as
\bea
t(s,\cos \theta)\al =\al
M_{D_1}^2+M_{D_2}^2-\frac{\left(s+M_{D_1}^2-M_{\phi_1}^2\right)\left(s+M_{D_2}
^2-M_ { \phi_2 } ^2\right) } {2s} \nonumber \\
\al -\al \frac{\cos \theta}{2s}\sqrt{\lambda(s,M_{D_1}^2,M_{\phi_1}^2)\lambda(s,M_{D_2}^2,M_{\phi_2}^2)},
\eea
where $\lambda(a,b,c)=a^2+b^2+c^2-2ab-2ac-2bc$ is the K{\"a}ll{\'e}n function.
We only deal with the $S$-wave scattering in this paper, and will drop the
subscript $l=0$ for brevity.

The unitarized two-body scattering amplitude has the
form~\cite{Oller:2000fj}
\bea
{T}(s)=[1-N(s)\cdot G(s)]^{-1}\cdot N(s), \label{eq:uni}
\eea
where the function $G(s)$ encodes the two-body right-hand cut and is given by the 
two-point loop function
\bea
G(s)=i\int
\frac{d^4q}{(2\pi)^4}\frac{1}{(q^2-M_D^2+i\epsilon)
\left[(q+p)^2-M_\phi^2+i\epsilon\right]}, \qquad s\equiv p^2\ .
\eea
The explicit expression for $G(s)$ reads~\cite{Oller:1998zr}
\bea
G(s)\al =\al \frac{1}{16\pi^2}\Big\{ a(\mu) + \ln \frac{M_D^2}{\mu^2}+\frac{s-M_D^2+M_\phi^2}{2s} \ln \frac{M_\phi^2}{M_D^2}\nonumber \\
\al \al +\frac{\sigma (s)}{2s}\Big[ \ln \big(\sigma (s)+s+M_D^2-M_\phi^2\big) - \ln \big( \sigma (s)-s-M_D^2+M_\phi^2\big) \nonumber \\
\al \al +\ln \big(\sigma(s)+s-M_D^2+M_\phi^2\big)-\ln \big( \sigma (s)-s+M_D^2-M_\phi^2\big) \Big]\Big\}, \label{eq:Gs}
\eea
where $a(\mu)$ is a subtraction constant with $\mu$ the renormalization scale
and $\sigma (s)=\sqrt{[s-(M_D+M_\phi)^2][s-(M_D-M_\phi)^2]}$.
Note that the logarithmic scale dependence can be absorbed into the subtraction
constance $a(\mu)$ and we do not distinguish the scale $\mu$ with the one introduced by dimensional regularization in the perturbative one-loop potentials.

While the right-hand cut effect is collected in the $G(s)$ function, the $N(s)$
function is free of any two-body right-hand cut. However,
it may include the left-hand cuts due to the crossed channels. Up to NNLO, the $N(s)$
function can be expressed as~\cite{Oller:2000fj,Guo:2011pa}
\bea
N(s)=\mathcal{V}_{\text{LO}}^\text{(WT+EX)}(s)+\mathcal{V}_{\text{NLO}}^\text{(CT)}(s)+\mathcal{V}_{\text{NNLO}}^\text{(CT+Loop)}(s)
-\mathcal{V}_{\text{LO}}^\text{(WT+EX)}(s)\cdot G(s)\cdot 
\mathcal{V}_{\text{LO}}^\text{(WT+EX)}(s). \label{eq:sub}
\eea
Eq.~(\ref{eq:uni}) is an algebraic approximation of the standard $N/D$
method\cite{Oller:1998zr}, and it should be understood
in the matrix form for coupled-channels, for which $G(s)= \text{diag}\{ G_i(s) \}$, with $i$ the channel index.

\section{Numerical analyses}\label{sec:num}

\subsection{Fit to lattice data of the scattering lengths}

Up to now, there is no experimental measurement on the light pseudoscalar
mesons scattering off heavy bosons. We can only rely on lattice QCD
results~\cite{Liu:2012zya,Mohler:2012na,Mohler:2013rwa,Lang:2014yfa,
Moir:2016srx} to determine the relevant LECs.
We will fit to the lattice results on the scattering lengths.
For the channels with definite strangeness and isospin, the $S$-wave scattering
lengths are obtained from the unitarized amplitudes ${T}(s)$ via
\bea
a^{(S,I)}_{D\phi \to D\phi}=-\frac{1}{8\pi (M_D+M_\phi) }{T}^{(S,I)}_{l=0}(s_{\text{thr}})_{D\phi \to D\phi},~~~s_\text{thr}=(M_D+M_\phi)^2.
\eea

Since the current lattice simulations are performed at unphysical pion masses
with fixed charm and strange quark masses,  in order to fit these lattice data,
one needs to know the pion mass dependence of the scattering lengths, which is
achieved by employing the following mass extrapolation
formulae:~\cite{Liu:2012zya}
\bea
M_K  = \sqrt{\mathring{M}_K^2+M_\pi^2/2 }, \quad
M_D  = \mathring{M}_D+(h_1+2h_0)\frac{M_\pi^2}{\mathring{M}_D} , \quad
M_{D_s}  = \mathring{M}_{D_s}+2h_0\frac{M_\pi^2}{\mathring{M}_{D_s}}, \label{massExtrapolation}
\eea
where $\mathring{M}_K$, $\mathring{M}_D$ and $\mathring{M}_{D_s}$ denote the masses in 
the two-flavor chiral limit
($M_\pi^2(\propto \hat{m})\to 0$ but with the fixed strange quark mass  $m_s$).
They have the form 
\bea
\mathring{M}_D^2=M_0^2+4h_0\mathring{M}_K^2,\quad
\mathring{M}_{D_s}^2=M_0^2+4(h_0+h_1)\mathring{M}_K^2.
\label{massring}
\eea
Using  Eqs.~(\ref{massExtrapolation}) and (\ref{massring}), one gets
\bea
h_1=\frac{M_{D_s}^2-M_D^2}{4(M_K^2-M_\pi^2 )},\label{h1value}
\eea
which is fixed as $h_1=0.427$ with the physical masses, i.e.,
$M_\pi^\text{Phy}=0.138~\text{GeV}$, $M_K^\text{Phy}=0.496~\text{GeV}$,
 $M_D^\text{Phy}=1.867~\text{GeV}$ and $M_{D_s}^\text{Phy}=1.968~\text{GeV}$.
Similar to the case of the pseudoscalar charmed mesons, the pion mass dependence of the 
masses of the vector mesons read,
consistent with the general expression derived in Ref.~\cite{Bruns:2004tj},
\bea
M_{D^*} = \mathring{M}_{D^*}+( \tilde{h}_1+2\tilde{h}_0)\frac{M_\pi^2}{\mathring{M}_{D^*}}, \quad
M_{D_s^*} = \mathring{M}_{D_s^*}+2\tilde{h}_0\frac{M_\pi^2}{\mathring{M}_{D_s^*}}.
\eea
Here, $\mathring{M}_{D^*}$ and $\mathring{M}_{D_s^*}$ denote the corresponding
two-flavor chiral limit masses,
which can be estimated by the relations
\bea\label{mdstar}
\mathring{M}_{D^*}-\mathring{M}_D \simeq M_{D^*}^\text{Phy}-M_D^\text{Phy}, \quad
\mathring{M}_{D_s^*}-\mathring{M}_{D_s} \simeq M_{D_s^*}^\text{Phy}-M_{D_s}^\text{Phy},
\eea
with $M_{D^*}^\text{Phy}$ and $M_{D_s^*}^\text{Phy}$ denoting the corresponding
physical masses, $2.008~\text{GeV}$
 and $2.112~\text{GeV}$, respectively.
One has $\tilde{h}_1=h_1$
and $\tilde{h}_0=h_0$ in the heavy
quark limit.\footnote{Analogous to Eq.~\eqref{h1value},
it is easy to see that 
$\tilde{h}_1=\left(M_{D_s^\ast}^2-M_{D^\ast}^2\right)/\left[
4\left(M_K^2-M_\pi^2\right)\right] =0.472$ , 
 which is close to $h_1$ numerically.}  These relations as well 
as similar relations for other LECs will be
employed in order to reduce the number of parameters.
The $DD^*\pi$ axial coupling constant $g_0$ can be fixed by the decay width 
$\Gamma_{D^{*+}\to D^0\pi^+}$. As discussed in Refs.~\cite{Yao:2015qia}, one 
gets
$g=(1.113\pm 0.147) ~\text{GeV}$ for the renormalized coupling $g$, which
contains the bare constants $g_0$ and one-loop chiral corrections.
At the one-loop level, the pion decay constant has the form~\cite{Gasser:1984gg}
\bea
\frac{F_\pi}{F_0}=1-\mu_\pi-\frac{1}{2}\mu_K+
4L_4^r(\mu)\frac{2M_K^2+M_\pi^2}{F_0^2}+4L_5^r(\mu)\frac{M_\pi^2}{F_0^2},
\eea
with $\mu_\phi=\frac{M_\phi^2}{16\pi^2 F_0^2}\ln \frac{M_\phi^2}{\mu^2}$.

In this paper, the scattering length data we use are taken from
Ref.~\cite{Liu:2012zya}  and
Refs.~\cite{Mohler:2012na,Mohler:2013rwa,Lang:2014yfa}, respectively. The
lattice simulation data for $M_K$, $M_D$, $M_{D_s}$ and $F_\pi$, $F_K$ are taken
from Refs.~\cite{Liu:2012zya} and~\cite{WalkerLoud:2008bp}, which share the same
ensembles (M007, M010, M020 and M030) with Ref.~\cite{Liu:2012zya}.
The fit results are listed in the Table~3 in Ref.~\cite{Yao:2015qia}.  In
addition, we also include the $DK$ scattering length with
$(S,I)=(1,0)$ at $M_\pi=0.156~\text{GeV}$~\cite{Mohler:2013rwa}, as discussed
in Ref.~\cite{Yao:2015qia}. It should be noticed that different lattice
configurations usually take
different values for both the strange and charm quark masses, which leads to
different values for $\mathring{M}_K$, $\mathring{M}_D$ and
$\mathring{M}_{D_s}$, as listed in Table~\ref{Tabmatrhing}.

\begin{table}[bt]
\caption{Parameters for the chiral extrapolation for different configurations. $h_0$, $L_4^r$ and $L_5^r$ are fixed by the data in Ref.~\cite{Liu:2012zya}.
The masses and decay constant in the chiral limit
are in units of GeV. $h_0$, $h_1$, $L_4^r$ and $L_5^r$ are dimensionless. The asterisk indicates an
input value.  }\label{Tabmatrhing}
\vspace{-0.5cm}
\bea
\begin{array}{c|cccccccccc}
\hline\hline
 & \mathring{M}_K &\mathring{M}_D    &\mathring{M}_{D_s}&F_0&h_{0}&h_1&10^{5}\cdot L_4^r& 10^{3}\cdot
 L_5^r\\
\hline
\text{Ref.~\cite{Liu:2012zya}} & 0.560 &1.940
&2.061&0.0733&0.0172&0.427^\ast & 0.951&1.326\\
\hline
\text{Ref.~\cite{Mohler:2013rwa}}& 0.486
&1.862&1.968&0.0762&0.0172&0.427^\ast & 0.951&1.326\\
\hline\hline
\end{array}\nonumber
\eea
\end{table}

Furthermore, in order to reduce the correlations between the LECs, we introduce
the following redefinitions of the LECs~\cite{Liu:2012zya,Yao:2015qia}
\bea
h_4^\prime \al=\al h_4 \bar{M}_D^2, \quad h_5^\prime =h_5 \bar{M}_D^2, \quad h_{24}=h_2+h_4^\prime , \quad  h_{35}=h_3+2h_5^\prime, \nonumber \\
g_1^\prime\al=\al g_1 \bar{M}_D,\quad g_2^\prime=g_2\bar{M}_D, \quad g_3^\prime=g_3\bar{M}_D^3, \quad   g_{23}=g_2^\prime-2g_3^\prime
\eea
where $\bar{M}_D$ stands for the average of the physical masses of the charmed mesons $D$ and $D_s$,
$\bar{M}_D=(M_D^\text{Phy}+M_{D_s}^\text{Phy})/2$. The new parameters $h_4^\prime$, $h_5^\prime$, $h_{24}$ and $h_{35}$ are dimensionless, and
$g_1^\prime$, $g_3^\prime$ and $g_{23}$ have the dimension of inverse mass. They are fixed by fitting to the lattice scattering lengths at varying
pion masses. It is well known that the state $D_{s0}^*(2317)$ in the
$(S,I)=(1,0)$ channel is produced as a bound state pole below the $DK$
threshold~\cite{Liu:2012zya,Guo:2015dha,Guo:2009ct}. Due to the large number of
the parameters and the small number of data, we constrain further the
subtraction constant $a(\mu)$ by requiring the existence of a bound state pole
at $2.317~\text{GeV}$ in the $(S,I)=(1,0)$ amplitude when all the parameters
take their physical values~\cite{Liu:2012zya}.

To compare with the result of Ref.~\cite{Yao:2015qia} 	where the $D^*$ mesons
are not included, we utilize the same fit procedures. In the fit UChPT-6(a), we
fit all of the data in Ref.~\cite{Liu:2012zya}, including 5 channels at
pion masses $0.301~\text{GeV}$, $0.364~\text{GeV}$, $0.511~\text{GeV}$ and $0.617~\text{GeV}$, as well as the isoscalar $DK$ channel at the pion mass
$0.156~\text{GeV}$ in Ref.~\cite{Mohler:2013rwa}, because all these refer to $N_f = 3$. 
However, as we know, the standard ChPT only works well in the small pion mass and low energy region. It
is naively expected
that the unitarized approach has a larger convergence range, but the convergence for the pion mass larger than $0.6~\text{GeV}$ is still questionable.
Therefore, to compare with the fit UChPT-6(a), another fit denoted as UChPT-6(b)
is performed excluding the lattice data at
$M_\pi=0.617~\text{GeV}$. Results of both fits are listed in Table~\ref{TabLECs}.

In both fits, as in  Ref.~\cite{Yao:2015qia}, the absolute value
of the dimensionless LEC $h_5^\prime$ is much larger than 1, too large to be
natural.
We therefore use the same method as therein to perform further fits by
minimizing the augmented $\chi^2$
\bea
\chi_{\rm aug}^2=\chi^2+\chi_{\rm prior}^2,
\eea
where $\chi^2$ is the standard chi-squared, and $\chi_{\rm prior}^2$ is a prior
 quantity constraining the LECs to take natural values. It is set to be the sum
 of squares of the free LECs . Two fits UChPT-6($a^\prime$) and
UChPT-6($b^\prime$) are obtained by minimizing $\chi_{\rm aug}^2$ instead of
$\chi^2$ using the
same data as in  UChPT-6(a) and UChPT-6(b). Although the values of LECs are more
natural, the $\chi^2$ values, with the prior parts subtracted, become very
large. As a result, the scattering lengths from
the new fits have larger deviations from the lattice data. More details about
the fit procedures can be found in Ref.~\cite{Yao:2015qia}.

Compared to the NLO fits~\cite{Guo:2015dha,Guo:2009ct}, the NNLO fits have
larger $\chi^2$ values, even though three more LECs $g_i\, (i=1,2,3)$ are 
included in the fits. This could be because the unitarization method we use
works better for the tree-level potentials than one-loop ones.
On the one hand, the left-hand cuts, stemming from the $t$- and $u$- channels, appear in
the one-loop potentials, which would cause the problem of violation of right-hand unitarity
in the region where the left- and right- cuts overlap, see more discussions in the next
section. On the other hand, the off-shell effects are partially included in the unitarized
amplitudes if the one-loop potentials are employed. Both of the above-mentioned effects have
non-trivial analytical structures and could make the NNLO unitarization much 
more cumbersome than the
NLO one.
Related to this is the fact that the scattering length $a_{D_s\pi\to 
D_s\pi}^{(1,1)}$ remains sizeable  in 
the $SU(2)$ chiral limit of $M_\pi \to 0$, as shown in Figs.~\ref{figUChPT5} and~\ref{figUChPT5prime}.
This was also the case in 
Ref.~\cite{Yao:2015qia}.\footnote{It is due to the nonvanishing $\mathring{M}_K$ 
in loops contributing to the $D_s\pi\to D_s\pi$ potential in the $SU(2)$ chiral 
limit. 
Near the $D_s\pi$ threshold, the $u$- and $t$-channel loops dominate in the 
$SU(2)$ chiral limit, which indicates that the left-hand cuts are 
non-negligible.  In the NLO case, the scattering length 
$a_{D_s\pi\to D_s\pi}^{(1,1)}$ is negligible in the $SU(2)$ chiral limit 
\cite{Liu:2012zya,Guo:2015dha}.
}

\begin{table}[tbh]
	\caption{Values of the LECs from the 6-channel fits using the method of
	UChPT. 	The $h_i$'s are dimensionless, and the  $g_1'$,
	$g_{23}$ and $g_3'$ are in GeV$^{-1}$.  }
	\label{TabLECs}
	\vspace{-0.5cm}
	\bea
	{
		\begin{array}{crrrr}
			\hline\hline
			&\text{UChPT-6(a)}&\text{UChPT-6(b)}&\text{UChPT-6($a^\prime$)}&\text{UChPT-6($b^\prime$)}\\
			&	\text{no prior}		 &	\text{no prior}			&\text{with prior}&\text{with
			prior}\\
			\hline
			h_{24}		& 0.44_{-0.07}^{+0.07}	& 0.49_{-0.08}^{+0.08}	& 0.52_{-0.09}^{+0.09}  & 0.61_{-0.10}^{+0.10} \\
			h_{35}		& 0.49_{-0.57}^{+0.68} 	& 1.03_{-0.91}^{+1.20}	& -0.19_{-0.22}^{+0.23} &  0.27_{-0.26}^{+0.27}\\
			h_4^\prime 	& -0.06_{-0.46}^{+0.48}	& -0.66_{-0.54}^{+0.54}	& -0.31_{-0.53}^{+0.55}	& -1.07_{-0.60}^{+0.60}\\
			h_5^\prime 	& -20.23_{-3.53}^{+3.04}&-23.91_{-8.98}^{+6.83} & -6.33_{-0.67}^{+0.66} & -3.68_{-0.76}^{+0.75}\\
			g_{1}^\prime& -2.17_{-0.32}^{+0.27} & -2.79_{-2.53}^{+0.55}	& -1.56_{-0.14}^{+0.12}	& -1.74_{-0.20}^{+0.16}\\
			g_{23} 		& -1.83_{-0.25}^{+0.21} & -2.33_{-0.49}^{+0.44} & -1.28_{-0.15}^{+0.14} & -1.38_{-0.21}^{+0.17}\\
			g_3^\prime 	& 3.20_{-0.57}^{+0.67} 	& 3.83_{-1.31}^{+1.71}	& 0.92_{-0.14}^{+0.14} 	& 0.19_{-0.18}^{+0.18} \\
			\hline
			\chi^2/{\rm d.o.f.}	&\frac{43.81}{21-7}=3.13 &\frac{14.26}{16-7}=1.58&\frac{143.78-45.36}{21-7}=7.03 &\frac{69.95-20.08}{16-7}=5.54	\\
			\hline
			\hline
		\end{array}\nonumber
	}
	\eea
\end{table}

Among various fits, UChPT-6(b) has the smallest $\chi^2$, which is also true for
the previous fits without $D^*$~\cite{Yao:2015qia}. In addition, the fits with
and without dynamical $D^*$ have similar values of the chi-squared and the LECs,
which indicates that  the influence of the $D^\ast$ on the quantities in
question is small.

\begin{figure}[htbp]
\begin{center}
\includegraphics[width=0.9\textwidth]{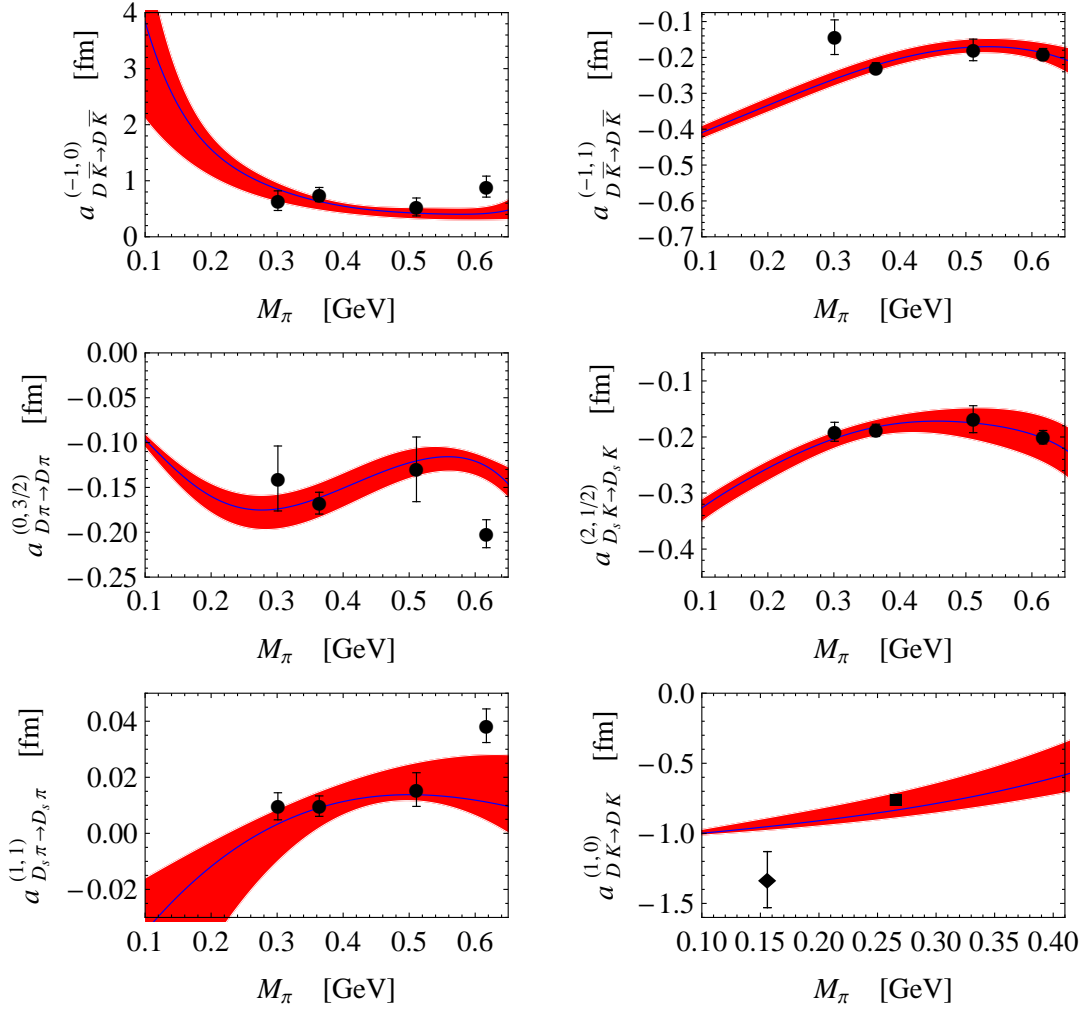}
\end{center}
\caption{The results of the UChPT-6(b) fits to the lattice data of the scattering lengths. The filled circles are lattice results in
Ref.~\cite{Liu:2012zya}, and the filled square (not included in the fits because it refers to $N_f=2$) 
and diamond are taken from Ref.~\cite{Mohler:2013rwa}. }\label{figUChPT5}
\end{figure}

\begin{figure}[htbp]
\begin{center}
\includegraphics[width=0.9\textwidth]{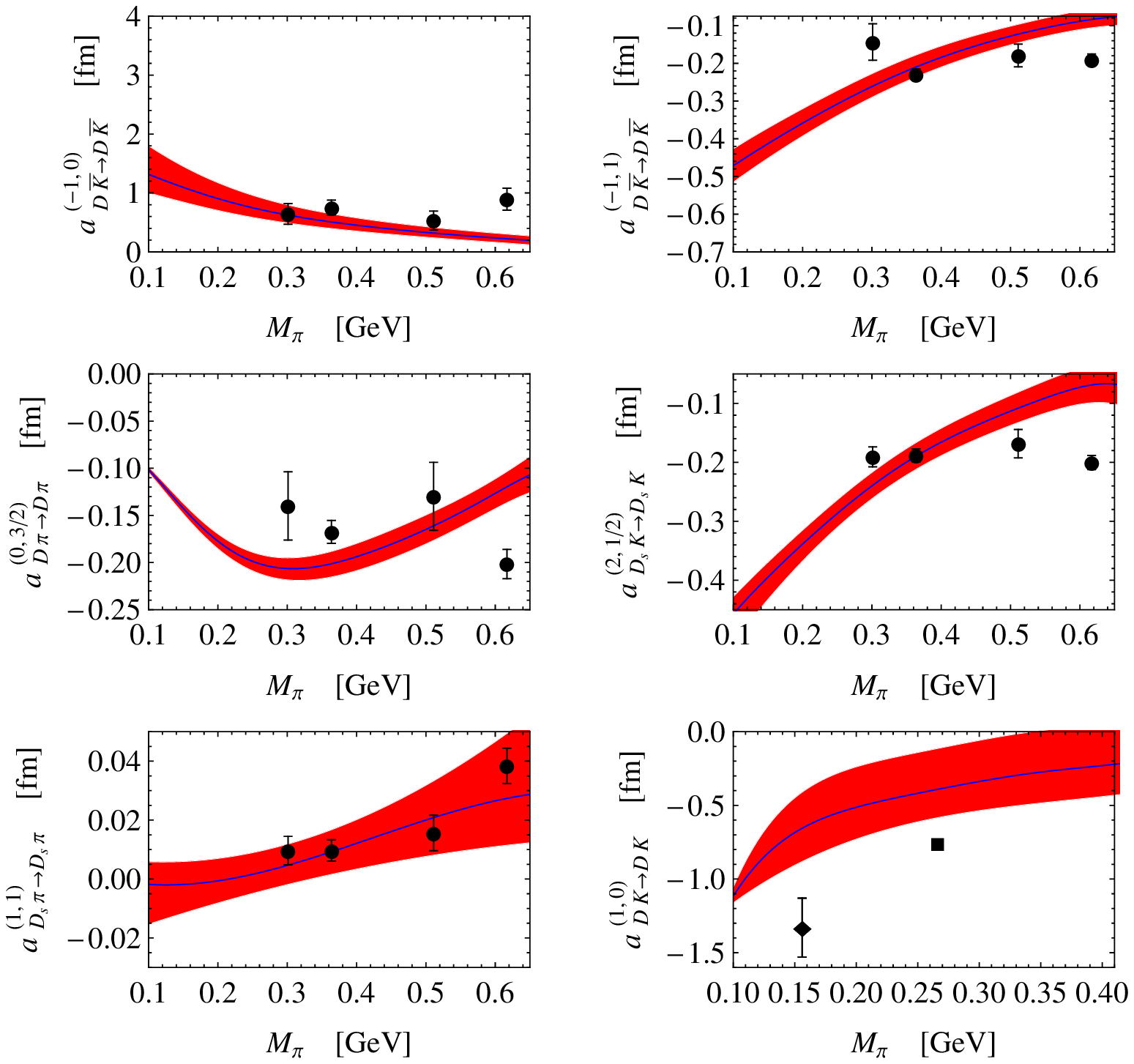}
\end{center}
\caption{The results of the UChPT-6($b^\prime$) fits to the lattice data of the scattering lengths. The filled circles are lattice results in
Ref.~\cite{Liu:2012zya}, and the filled square (not included in the fits because it refers to $N_f=2$) 
and diamond are taken from Ref.~\cite{Mohler:2013rwa}. }\label{figUChPT5prime}
\end{figure}

\subsection{Dynamically generated resonances}

The unitary $S$-matrix could have poles in the complex energy ($\sqrt{s}$) plane
in the region not far from the relevant thresholds.  Bound states and resonances
are poles located on the physical and unphysical Riemann sheets, respectively.
Different Riemann sheets are characterized by the sign of the imaginary part of the
loop function on the right branch cuts. Each loop function $G_i(s)$ has two
sheets: the physical/first Riemann sheet and the unphysical/second Riemann
sheet, denoted as $G^i_I(s)$ and $G^i_{II}(s)$, respectively.
The expression in Eq.~(\ref{eq:Gs}) defines the physical Riemann sheet, while
the expression on the second sheet is given by analytic continuation
via~\cite{Oller:1997ti}
\bea
G^i_{II}(s+i\epsilon)=G^i_I(s+i\epsilon)-2i~\text{Im}~G^i_I(s+i\epsilon).
\eea
For the $n$-channel case, there exist $2^n$ Riemann sheets in total. Different
sheets can be accessed by properly choosing the loop functions $G^i_{I/II}(s)$.
We use the sign of the imaginary
part of $G^i(s)$ above threshold to indicate the $G^i_{I/II}(s)$. In this
convention,  for the coupled-channel case, the first Riemann sheet is labelled
as $(+,+,+,\ldots)$, while $(-,+,+,\ldots)$, $(-,-,+,\ldots)$, $(-,-,-,\ldots)$
and so on correspond to the second, third, fourth, \ldots sheets, respectively.
Normally, at a given energy $s$, only the sheet which can be reached from the
physical one by crossing the branch cut from $s+i\epsilon$ to $s-i\epsilon$
between the thresholds $\text{thr}_{n-1}$ and $\text{thr}_n$,
has a significant impact on physical observables.

As discussed earlier, we have found that the impact of the $D^*$ mesons on the
$S$-wave $D\phi$ scattering processes is very small. We thus search for poles
using the amplitudes without $D^*$ derived in Ref.~\cite{Yao:2015qia}. In this
way, the complexity of analytically continuing the three- and four-point loops
in Fig.~\ref{fig:FD} is avoided. The physical meson masses and decay constant
are employed in the pole searching, and the obtained poles are listed in
Table.~\ref{Tabpolec}.

\begin{figure}[htbp]
\begin{center}
\includegraphics[width=0.84\textwidth]{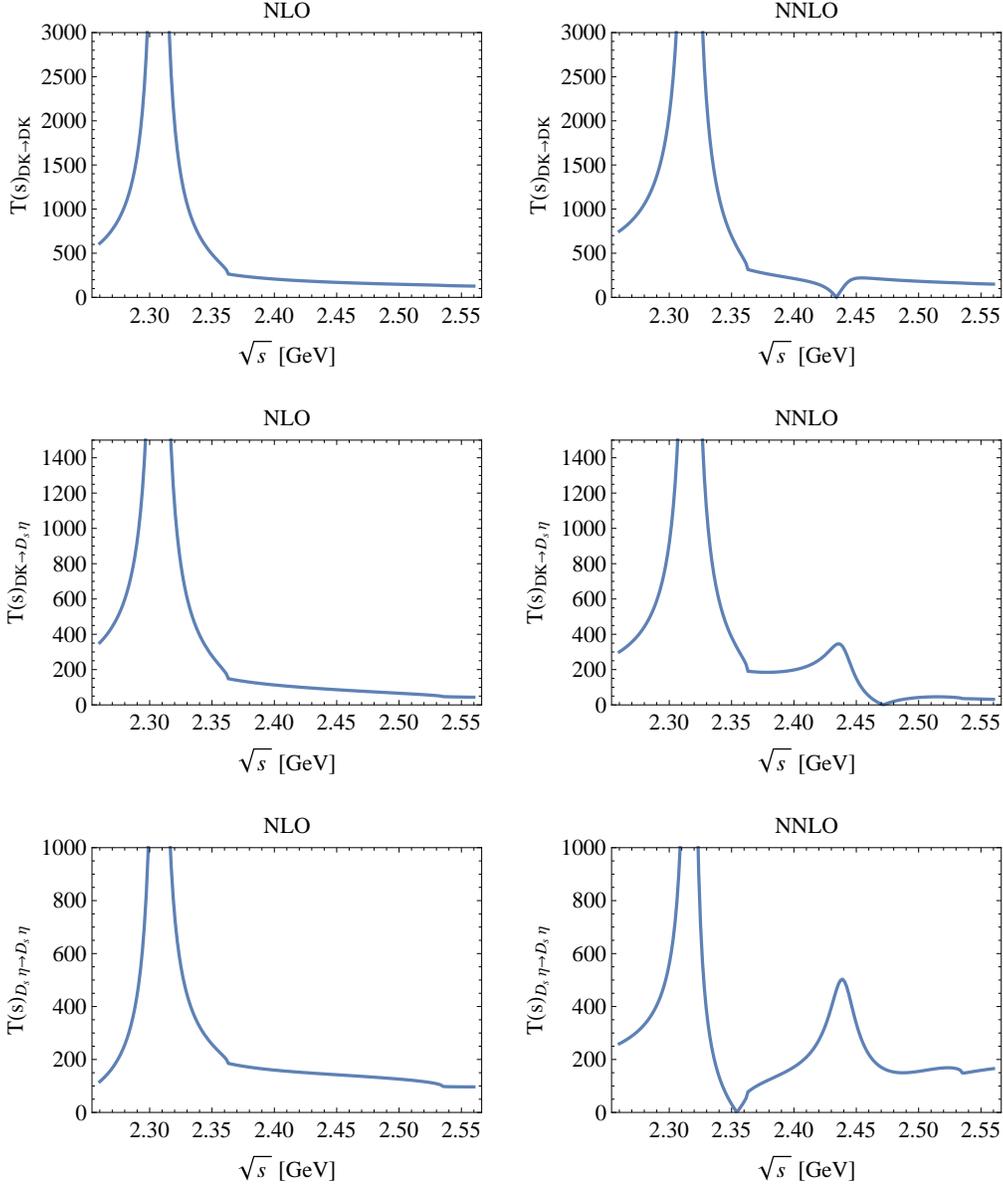}
\end{center}
\caption{Absolute values of amplitudes for $(S,I)=(1,0)$
in NLO and NNLO calculations, respectively.}\label{fig:Dseta}
\end{figure}
For the $(S,I)=(1,0)$ coupled-channel system, in addition to the pole at
$\sqrt{s}=2.317$~GeV on the physical sheet, which corresponds to $D_{s0}^\ast
(2317)$ and was used as a condition to constrain the parameters, using the
central values of the parameters we also found
a pair of poles with a small but nonvanishing imaginary part on the second
Riemann
sheet, $\sqrt{s}=(2.439\pm i0.01)$~GeV. The only work which reported an
analogous pole is Ref.~\cite{Guo:2015dha}, where a virtual state at
$\sqrt{s}=2.356$~GeV below $DK$ threshold on the second Riemann sheet was
reported in an NLO calculation including $DK,D_s\eta$ and $D_s\eta'$ channels.
We check whether such a pole exists using the parameters from the NLO fits in
Ref.~\cite{Liu:2012zya}, and found that only part of the allowed parameter space
allows for the pole on the unphysical Riemann sheet. Moreover, the effect of
this virtual state pole located at $\sqrt{s}=2.356$~GeV on the physical
amplitude is negligible in the NLO calculation, as can be seen from the left
column of Fig.~\ref{fig:Dseta}. However, the pole on the second Riemann sheet in
the NNLO calculation can have a non-negligible effect on specific physical
amplitudes, as shown in the right column of Fig.~\ref{fig:Dseta}. These different
behaviors are mainly due to different locations of the poles. Nevertheless, we
see that the lattice data on the scattering lengths are insufficient to
constrain the parameters, and as a result, calculations at different orders may
even have a sizeable discrepancy in amplitudes not far from thresholds.  More
lattice data on $D\phi$ scattering observables are needed to better pin
down the LECs.

In addition, we also found a pair of poles $\sqrt{s}=(2.534\pm i0.097)$ GeV
on the second Riemann sheet which are not included in Table~\ref{Tabpolec}. They
have a negligible effect on physical amplitudes and would disappear if the $u$-
and $t$-channels are turned off.
Likewise, we do not include the following poles in Table~\ref{Tabpolec} since
they are located far from the physical region and have little effect: poles
at $\sqrt{s}=(2.448\pm i0.049)$~GeV and $(2.267\pm i0.099)$~GeV on the third
Riemann sheet for $(S,I)=(1,0)$ and $(1,1)$, respectively; poles at
$(2.257\pm i0.018)$~GeV on the second Riemann sheet in the $(-1,0)$
$D\bar{K}\to D\bar{K} $ channel.

It is well-known that the unitarization approach, relying on right-hand 
unitarity and the on-shell approximation,  has the problem of violation of unitarity 
when the left-hand cut occurs in the on-shell potential. For instance, the left-hand 
cut in the $K\bar K\to K\bar K$ amplitude leads a violation of unitarity for 
the $\pi\pi$ scattering in the $\pi\pi$--$K\bar K$ coupled-channel 
system~\cite{GomezNicola:2001as,Dai:2011bs}.~\footnote{As pointed out by Refs~\cite{Guerrero:1998ei, GomezNicola:2001as}, 
the unitarity violation is numerically small in the $\pi\pi$--$K\bar K$ case, 
hence no serious problem was caused there.} 
The same unitarity violation happens to the $D\phi$ scattering with $(S,I)=(0,1/2)$, 
which has three coupled channels: $D\pi$, $D\eta$ and $D_s\bar K$. 
One of the left-hand cuts from the inelastic
channel $ D_s\bar{K}\to D\eta$ amplitude, from ($1.488$ GeV)$^2$ to ($2.318$ 
GeV)$^2$,
overlaps with the right-hand cut starting from the $D\pi$ threshold, which can
be verified by the discontinuity across the real axis below the $D\pi$
threshold. Although this left-hand cut is not numerically important, its 
presence together with other left-hand cuts and right-hand cuts make the 
whole real axis nonanalytic. 
Since Eq.~\eqref{eq:uni} was derived using the $N/D$ method 
neglecting the left-hand cuts, its continuation to the complex plane 
near the left-hand cut is untrustworthy.
As a result, the 
coupled-channel amplitudes obtained from Eq.~\eqref{eq:uni} do not have the 
correct analytic properties even in the relevant energy region. Consequently, 
a pair of pole at $(2.046\pm i 0.050)~\text{GeV}$ are found on the first Riemann 
sheet  for the coupled-channel $(S,I)=(0,1/2)$ amplitude. As we know, poles on 
the first Riemann sheet can only be located on the real axis below the lowest 
threshold, which are associated with bound states. A pole on the first sheet 
with a nonvanishing imaginary part or above the lowest threshold is inconsistent 
with causality. The appearance of the pole on the first sheet in the coupled-channel 
$(S,I)=(0,1/2)$ is due to the existence of the coupled-channel cut. The 
left-hand cuts stem from the one-loop potentials, and are absent in the NLO cases.

If we consider only the single-channel $D\pi$ for $(S,I)=(0,1/2)$, there is no
such a problem as it comes from the left-hand cut of the inelastic channels.
We searched for poles in the single-channel amplitude, and found a pair of poles
in the second Riemann sheet given in Table~\ref{Tabpoles},\footnote{Notice that
the poles found in both the single-channel and coupled-channel unitarized NLO
amplitudes are similar to each other in Ref.~\cite{Guo:2009ct}.}
corresponding to the lower pole at $(2.105- i0.102)$~GeV of the two-pole
structure of $D_0^*(2400)$ advocated in
Ref.~\cite{Albaladejo:2016lbb}.

\begin{table}[t]
\caption{Poles in the coupled-channel amplitudes based on UChPT-6(b) in Table 4 of
Ref.~\cite{Yao:2015qia}. Physical masses and decay constants are used to obtain
the poles.
The Riemann sheets on which the poles are located are indicated in the last
column.
}\label{Tabpolec}
\vspace{-0.5cm}
\bea
\begin{array}{lll|cc|c}
\hline\hline
(S,I)	& 		\text{Channel} 	& \text{Thr(MeV)}   & \text{Re(MeV)}	  & \text{Im(MeV)}	 &	 \text{RS}	 \\
\hline
( 1,0)  & DK\to DK			&  2363		   &  2317  & 		
0	 &	 \text{I}   \\
 \,		& D_s\eta \to D_s\eta	&  2535		   &  2439  &     \pm 10 &	 \text{II}  \\
 \hline
( 1,1)  & D_s\pi \to D_s\pi     &  2106 	   &  2378  &   \pm 19   &   \text{II}  \\
\,	    & DK \to DK				&  2363   &  \,	&	\,   &	 
\, \\
\hline\hline
\end{array}\nonumber
\eea
\end{table}	

\begin{table}[t]
\caption{Poles in the single-channel amplitudes based on UChPT-6(b) in Table 4 of
Ref.~\cite{Yao:2015qia}. Physical masses are used to obtain the poles.
The Riemann sheets on which the poles are located are indicated in the last
column.
}\label{Tabpoles}
\vspace{-0.5cm}
\bea
\begin{array}{lll|cc|c}
\hline\hline
(S,I)	& 	\text{Channel} 	& \text{Thr(MeV)}   & \text{Re(MeV)}	  & \text{Im(MeV)}	 &	 \text{RS}	 \\
\hline
( 1,0)  & DK\to DK				&  2363		   &  2277  & 		
0	 &	 \text{I}   \\
 \,		& \, 					&  \,		   &  2436  &    \pm 15  &   \text{II}  \\
 \hline
(0,1/2) & D\pi \to D\pi 	    &  2005 	   &  2107  &    \pm 82  &   \text{II}  \\
\hline\hline
\end{array}\nonumber
\eea
\end{table}

In addition, we also investigated the pole movements with varying pion masses.
The pion mass dependence trajectories of the poles can provide us with useful
information about the properties of the different states, as discussed, e.g., in
Ref.~\cite{Hanhart:2014ssa}. The $M_\pi$ trajectory for the pole corresponding
to $D_{s0}^*(2317)$ is plotted in Fig.~\ref{fig2317}. The pole positions on the
first Riemann sheet, which are identified as the pole mass, are shown as the
solid line. The dotted line stands for the trajectory of the $DK$ threshold.
From Fig.~\ref{fig2317}, one can see that the $D_{s0}^*(2317)$ always stays
below the corresponding $DK$ threshold as a bound state for a wide range of
$M_\pi$. The trajectory of $D_{s0}^*(2317)$ is quite similar to the NLO fit
result, as shown in Ref.~\cite{Guo:2015dha}.

\begin{figure}[htbp]
\begin{center}
\includegraphics[width=0.78 \textwidth]{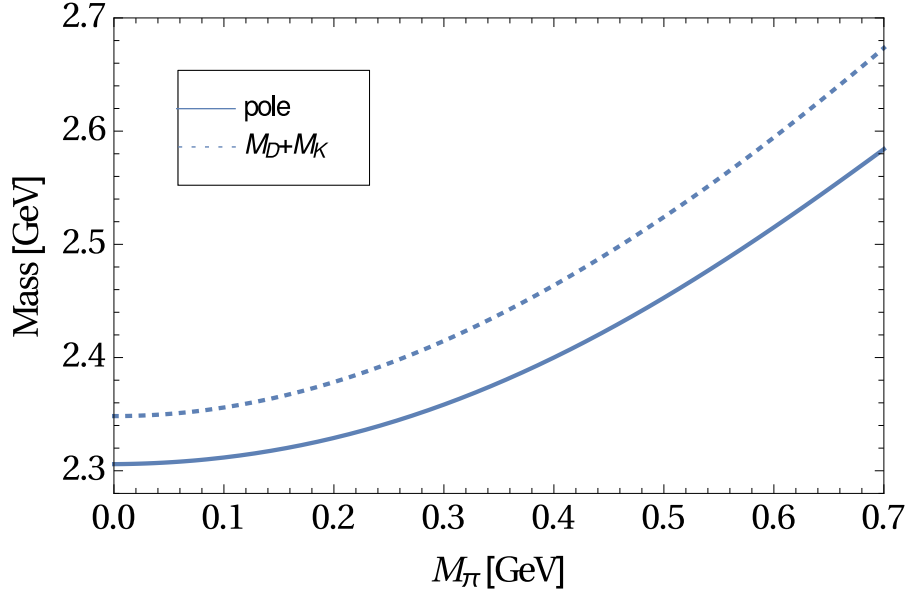}
\end{center}
\caption{The trajectory of the pole $D_{s0}^*(2317)$ with varying $M_\pi$. }\label{fig2317}
\end{figure}

\begin{figure}[htbp]
\begin{center}
\includegraphics[width=0.75 \textwidth]{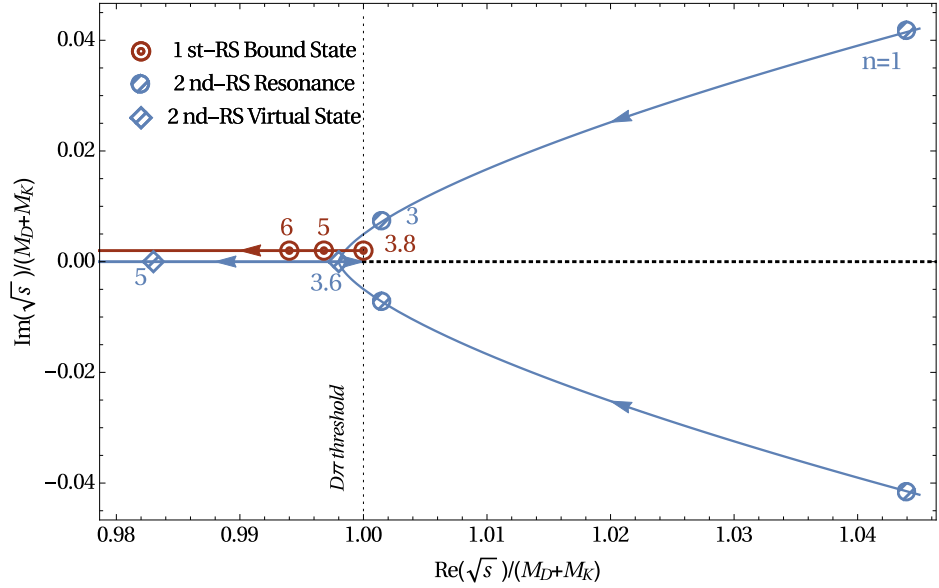}
\end{center}
\caption{The trajectory of the pole around $2.1~\text{GeV}$ for the single-channel 
$(S,I)=(0,1/2)$ with varying $M_\pi$. }\label{fig012}
\end{figure}

On the contrary, the pion mass dependence trajectory of the pole around
$2.1~\text{GeV}$ on the second Riemann sheet for the single-channel
$(S,I)=(0,1/2)$ in Table~\ref{Tabpoles} is quite complicated, as shown in
Fig.~\ref{fig012}. As the value of $M_\pi$ increases from $M_\pi^\text{Phy}$,
both the real and imaginary parts of the pole tend to decrease on the second
Riemann sheet. At some point around $3.2M_\pi^\text{Phy}$, the real part of the
pole becomes
 lower than the corresponding $D\pi$ threshold. When $M_\pi$ increases to around
 $3.6M_\pi^\text{Phy}$, the pair of poles hits 
the real axis below the threshold and these become two virtual states on the
second Riemann sheet. If we further increase the pion mass, one of the virtual
poles would move along the real axis away from the threshold, while the other
one moves towards the threshold and becomes a bound state of the first Riemann
sheet at $3.8M_\pi^\text{Phy}$. If we keep increasing $M_\pi$, both the virtual
state and bound state move away from the threshold along the real axis. The
behavior of the pole is similar to the corresponding ones in
Refs.~\cite{Guo:2009ct,Guo:2015dha} as well as the pion mass dependence of the
$f_0(500)$ in Ref.~\cite{Hanhart:2008mx}. It is general for $S$-wave states.

\section{Summary and conclusions} \label{sec:sum}

We have calculated the potential for the scattering of Goldstone bosons off
charmed mesons, including the charmed vector mesons as explicit degrees of freedom,
up to the NNLO in a framework of covariant ChPT. We explicitly show
that the UV divergences and the so-called power counting breaking terms from the
one-loop potentials can be absorbed by a redefinition of the LECs. In the EOMS
scheme, we obtained the $D\phi$ scattering potentials possessing the good
properties, i.e., they are free of UV divergences and power counting violating
terms. In order to describe the $S$-wave scattering lengths at high pion masses
and to study the possible dynamically generated resonances that are absent in the 
Lagrangian, e.g. the $D_{s0}^*(2317)$, at relatively high energies, which is the nonperturbative
effect, a unitarization procedure is employed.

In order to determine the LECs, we performed fits to scattering lengths in a few
channels computed in lattice QCD at various unphysical pion masses.
Since the lattice simulations are performed with fixed charm and strange quark
masses with varying up and down quark masses, we derived the corresponding pion
mass dependence of the scattering lengths by extrapolation of the involved
masses and the pion decay constant. For an easy comparison to the previous case
without including $D^*$ explicitly, we used  a similar fit procedure as in
Ref.~\cite{Yao:2015qia}. It turns out that, for UChPT-6(b), the current fit
result is quite close to the previous one which was done without explicit
vector charmed mesons.
It is thus a firm conclusion that the $D^*$ contribution to the $S$-wave $D\phi$
scattering in the threshold region is negligible.

Based on the small contribution of the $D^*$ to the $D\phi$ scattering
potentials, we investigated possible dynamically generated resonances using the
unitarized scattering amplitudes without explicit $D^*$ by analytic
continuation. It is worth noticing that a pair of poles with nonvanishing
imaginary parts are found on the physical Riemann sheet in the coupled-channel
$(S,I)=(0,1/2)$ amplitude, which are at odds with causality. The issue is caused
by the coupled-channel left-hand cut, which is not taken into account in the
unitarization procedure we used. It may be avoided by using a single-channel
potential if we only focus on the region near the $D\pi$ threshold.
In the end, we studied the trajectories of the poles corresponding to the
$D_{s0}^*(2317)$ and a resonance in $(S,I)=(0,1/2)$ channel with varying the
pion mass.
They exhibit similar behaviors as in  the NLO case given by
Ref.~\cite{Guo:2015dha}.
To summarize, the LECs are badly determined due to the 
scarcity of available data. Thus to come to firmer conclusions, more lattice data
are required, as also concluded from investigations based on the
resonance-exchange model and $S$-matrix properties in Ref.~\cite{Du:2016tgp}.

\section*{Acknowledgements}
We would like to thank Jiunn-Wei~Chen, Jambul~Gegelia and Zhi-Hui~Guo for helpful discussions.
MLD acknowledges the warm hospitality of the ITP of CAS where part of this work
was done.
This work is supported in part by DFG and NSFC through funds provided to the
Sino-German CRC 110 ``Symmetries and the Emergence of Structure in QCD" (NSFC
Grant No.~11621131001, DFG Grant No.~TRR110), by NSFC (Grant No.~11647601), by 
the Thousand Talents Plan for Young Professionals, by
the CAS Key Research Program of Frontier Sciences (Grant No.~QYZDB-SSW-SYS013),
and by the CAS President's International Fellowship Initiative (PIFI) (Grant
No.~2017VMA0025).

\bigskip

\appendix
\section{Renormalization of the LECs within EOMS scheme}
\label{sec:app}

In this appendix, we use the following notations for the chiral limit masses of $D$ and $D^\ast$: $m_D=M_0$ and $m_{D^\ast}=M_0^{\ast}$.  Following Ref.~\cite{Denner:1991kt}, the $N$-point one-loop integrals are defined by
\bea\label{loopintdef}
T^N&=& \frac{(2\pi\mu)^{4-d}}{i\pi^2}
\int \frac{{\rm d}^d k}{\left[k^2-m_{1}^2+i\epsilon\right]\left[(k+p_1)^2-m_{2}^2
+i\epsilon\right]\cdots \left[(k+p_{N-1})^2-m_{N}^2+i\epsilon\right]}.
\nonumber
\eea
The one-, two- and  three-point one-loop scalar integrals are denoted by $A$, $B$ and $C$ as follows:
\bea
T^1=A_0(m_1^2)\ ,\quad
T^2 =B_0(p_1^2,m_1^2,m_2^2)\ ,\quad
T^3=C_0(p_1^2,(p_1-p_2)^2,p_2^2,m_1^2,m_2^2,m_3^2)\, .
\eea

\subsection{$\beta$-functions\label{sec:beta}}
The $\beta$-functions in Eq.~(\ref{eq:UVshift}) read
\bea
\beta_{M_0^2} & = & -\frac{4\g^2 (3\md^2-\mx^2)}{9}\ ,\nonumber\\
\beta_{M_0^{\ast 2}} &=& \frac{4\g^2 \md^2(3\mx^2-\md^2)}{3\mx^2}\ ,\nonumber\\
\beta_{h_0}&=&\frac{11\,\g^2\,\md^2}{24\,\mx^2}\ ,\nonumber\\
\beta_{h_1}&=&\frac{5\,\g^2\,\md^2}{8\,\mx^2}\ ,\nonumber\\
\beta_{h_2}&=&\frac{\md^2(\mx^4-22\,\g^2\,\mx^2+4\,\g^4)}{48\,\mx^4}\ ,\nonumber\\
\beta_{h_3}&=&\frac{-9\,\md^2\mx^4+18\,\g^2(3\,\md^2\mx^2+16\mx^4)+4\g^4(\md^2+2\mx^2)}{144\,\mx^4}\ ,\nonumber\\
\beta_{h_4}&=&\frac{1}{24}\left(7-\frac{10\,\g^2}{\mx^2}+\frac{4\,\g^4}{\mx^4}\right)\ ,\nonumber\\
\beta_{h_5}&=&-\frac{7}{16}+\frac{9\,\g^2}{8\mx^2}-\frac{13\,\g^4}{18\mx^4}\ ,\nonumber\\
\beta_{g_0}&=&-\g\,\mx^2+\g^3\left(\frac{7}{4}-\frac{5\,\md^2}{4\,\mx^2}\right)\ ,\nonumber\\
\beta_{g_1}&=&\frac{-41\,\g^2\,\mx^2+30\,\g^4}{288\,\mx^4}\ ,\nonumber\\
\beta_{g_2}&=&-\frac{9}{128}+\frac{67\,\g^2}{288\,\mx^2}-\frac{3\,\g^4}{16\,\mx^4}\ ,\nonumber\\
\beta_{g_3}&=&0\ .
\eea
\newpage
\subsection{Coefficients of finite shifts\label{sec:finite}}
In this appendix, we express the EOMS subtractions in terms of the standard loop function.
The explicit coefficients of the finite shifts in Eq.~(\ref{eq:finiteshift}) are
\bea
\bar{\beta}_{h_0}&=&\frac{11 \md^2}{36 \mx^2}\g^2+\frac{11}{24 \mx^2}\g^2\ads-\frac{11 \left(\md^2+\mx^2\right)}{24 \mx^2}\g^2\bdd\ ,\\
\bar{\beta}_{h_1}&=&\frac{5 \md^2}{12 \mx^2}\g^2+\frac{5}{8 \mx^2}\g^2\ads-\frac{5 \left(\md^2+\mx^2\right)}{8 \mx^2}\g^2\bdd\ ,\\
\bar{\beta}_{h_2}&=&\left[-\frac{\md^2}{72}-\frac{1}{144} \left(\frac{31 \md^2}{\mx^2}+3\right)\g^2+\frac{\md^2 \left(\mx^2-7 \md^2\right)}{72 \mx^4 (\md-\mx) (\md+\mx)}\g^4\right]\nonumber\\
&+&\left[-\frac{1}{48}+\frac{3 \md^2-4 \mx^2}{24 \md^2 \mx^2-24 \mx^4}\g^2+\frac{-3 \md^4+6 \md^2 \mx^2-4 \mx^4}{12 \mx^4 (\md-\mx)^2 (\md+\mx)^2}\g^4\right]\add\nonumber\\
&+&\left[\frac{-8 \md^4+8 \md^2 \mx^2+\mx^4}{24 \md^2 \mx^2 \left(\md^2-\mx^2\right)}\g^2+\frac{-2 \md^4+5 \md^2 \mx^2-2 \mx^4}{12 \mx^4 \left(\md^2-\mx^2\right)^2}\g^4\right]\ads\nonumber\\
&+&\left[\frac{1}{24} \left(\frac{8 \md^2}{\mx^2}+\frac{\mx^2}{\md^2}+9\right)\g^2+\frac{\md^2+\mx^2}{6 \mx^4}\g^4\right]\bdd\ ,\\
\bar{\beta}_{h_3}&=&\left[\frac{\md^2}{24}+\left(\frac{3 \md^2}{4 \mx^2}+\frac{1}{6}\right)\g^2+\frac{7 \md^4-65 \md^2 \mx^2+112 \mx^4}{216 \mx^4 \left(\mx^2-\md^2\right)}\g^4\right]\nonumber\\
&+&\left[\frac{1}{16}+\frac{1}{4 \mx^2}\g^2+\frac{27 \md^4-46 \md^2 \mx^2+58 \mx^4}{36 \mx^4 (\md-\mx)^2 (\md+\mx)^2}\g^4\right]\add\nonumber\\
&+&\left[\frac{3}{8} \left(\frac{1}{\mx^2}-\frac{2}{\md^2}\right)\g^2-\frac{4 \md^6-2 \md^4 \mx^2+68 \md^2 \mx^4+8 \mx^6}{72 \mx^4 \left(\md^3-\md \mx^2\right)^2}\g^4\right]\ads\nonumber\\
&+&\left[\left(\frac{1}{4}-\frac{\md^2}{4 \mx^2}\right)\g^2+\frac{-15 \md^6+29 \md^4 \mx^2-\md^2 \mx^4+35 \mx^6}{18 \mx^4 \left(\md^2-\mx^2\right)^2}\g^4\right]\bds\nonumber\\
&+&\left[-\frac{3}{8} \left(\frac{\md^2}{\mx^2}-\frac{2 \mx^2}{\md^2}+7\right)\g^2+
\left.\frac{\md^8-5 \md^6 \mx^2-27 \md^4 \mx^4-19 \md^2 \mx^6+2 \mx^8}{18 \mx^4 \left(\md^3-\md \mx^2\right)^2}\g^4\right] \right. \nonumber\\
& & \bdd \nonumber\\
&+& \left[\frac{\left(\md^2+3 \mx^2\right)^2}{3 \mx^2 \left(\mx^2-\md^2\right)}\g^4 \right] \cds\,\\
\bar{\beta}_{h_4}&=&\left[-\frac{35}{72}+\frac{1}{36} \left(\frac{3}{\md^2}+\frac{1}{\mx^2}\right)\g^2+\frac{2 \md^2-8 \mx^2}{9 \md^2 \mx^4-9 \mx^6}\g^4\right]\nonumber\\
&+&\left[-\frac{7}{24 \md^2}+\frac{2}{3 \md^4-3 \md^2 \mx^2}\g^2-\frac{2}{3 \md^2 (\md^2-\mx^2)^2 }\g^4\right]\add\nonumber\\
&+&\left[-\frac{5 \md^4+\md^2 \mx^2+2 \mx^4}{12 \md^6 \mx^2-12 \md^4 \mx^4}\g^2+\frac{\left(\md^2+\mx^2\right)^2}{6 \mx^4 \left(\md^3-\md \mx^2\right)^2}\g^4\right]\ads\nonumber\\
&+&\left[\frac{1}{12} \left(-\frac{2 \mx^2}{\md^4}-\frac{3}{\md^2}+\frac{5}{\mx^2}\right)\g^2-\frac{\md^2+\mx^2}{6 \md^2 \mx^4}\g^4\right]\bdd\ ,
\eea
\bea
\bar{\beta}_{h_5}&=&\left[\frac{35}{48}-\frac{1}{4\mx^2}\g^2-\frac{31\md^2+3\mx^2}{54\md^2\mx^4}\g^4\right]+\left[\frac{7}{16\md^2}-\left(\frac{1}{8\mx^4}+\frac{1}{\md^2\mx^2}\right)\g^2\right.\nonumber\\
& & +\left. \frac{-23 \md^6+74 \md^4 \mx^2-95 \md^2 \mx^4+20 \mx^6}{36 \md^2 \mx^6 (\md-\mx)^2 (\md+\mx)^2}\g^4\right]\add\nonumber\\
&+&\left[\frac{\md^6+10 \md^4 \mx^2+17 \md^2 \mx^4-4 \mx^6}{36 \md^4 \mx^4 \left(\md^2-\mx^2\right)^2}\g^4\right]\ads\nonumber\\
&+&\left[\frac{5 \md^8+17 \md^6 \mx^2+39 \md^4 \mx^4-17 \md^2 \mx^6+4 \mx^8}{36 \md^4 \mx^4 \left(\md^2-\mx^2\right)^2}\g^4\right]\bdd\nonumber\\
&+&\left[\frac{\md^2-\mx^2}{8 \mx^4}\g^2+\frac{23 \md^6-53 \md^4 \mx^2+25 \md^2 \mx^4-43 \mx^6}{36 \mx^6 \left(\md^2-\mx^2\right)^2}\g^4\right]\bds\nonumber\\
&-&\left[\frac{\left(\md^2+3 \mx^2\right)^2}{6 \mx^4 \left(\mx^2-\md^2\right)}\g^4\right]\cds\ ,\\
\bar{\beta}_{g_0}&=&\left[\frac{1}{12} \left(3 \md^2-\mx^2\right)\g+\frac{1}{72} \left(41-\frac{39 \md^2}{\mx^2}\right)\g^3\right]\nonumber\\
&+&\left[\frac{\md^2+\mx^2}{8 \mx^2}\g+\frac{5 \md^4-7 \md^2 \mx^2+9 \mx^4}{12 \md^2 \mx^4-12 \mx^6}\g^3\right]\add\nonumber\\
&+&\left[\frac{3 \left(\md^2+\mx^2\right)}{8 \md^2}\g-\frac{17 \md^4-19 \md^2 \mx^2+9 \mx^4}{12 \md^4 \mx^2-12 \md^2 \mx^4}\g^3\right]\ads\nonumber\\
&-&\left[\frac{3 \left(\md^2-\mx^2\right)^2}{8 \md^2}\g-\frac{15 \md^6-33 \md^4 \mx^2-7 \md^2 \mx^4+9 \mx^6}{12 \md^4 \mx^2-12 \md^2 \mx^4}\g^3\right]\bdd\nonumber\\
&-&\left[\frac{\left(\md^2-\mx^2\right)^2}{8 \mx^2}\g+\frac{-5 \md^6+7 \md^4 \mx^2+5 \md^2 \mx^4+9 \mx^6}{12 \mx^4 \left(\mx^2-\md^2\right)}\g^3\right]\bds\nonumber\\
&-& \left[ \frac{\left(\md^2+3 \mx^2\right)^2}{6 \mx^2}\g^2\right]\cds\ .
\eea
Here the involved scalar one-loop integrals stand for
their finite parts only, which are obtained from the original ones, 
defined in Eq.~(\ref{loopintdef}), by performing the $\overline{MS}-1$ 
subtraction.

\end{document}